\documentclass{aa}

\usepackage{graphicx}
\usepackage{txfonts}
\usepackage{siunitx}

\usepackage{xcolor}
\usepackage[normalem]{ulem}
\usepackage{amsmath}
\usepackage{hyperref}
\hypersetup{
  colorlinks=true,
  linkcolor=blue,
  urlcolor=blue,
  citecolor=blue
  }

\DeclareSIUnit\erg{erg}
\DeclareSIUnit\dyne{dyn}
\DeclareSIUnit\dn{DN}
\DeclareSIUnit\maxwell{Mx}
\DeclareSIUnit\gramme{g}
\DeclareSIUnit\pix{g}
\DeclareSIUnit\angstrom{\text {Å}}

\begin{document}

\newcommand{\ad}{\textcolor{red}}
\newcommand{\software}{\textit}
\newcommand{\elsasser}{Els\"asser}
\newcommand{\alfven}{Alfv\'en}
\newcommand{\vs}{{v_\mathrm{s}}}

  \title{Fast and periodic propagating disturbances along coronal loops detected with EUI on board Solar Orbiter}

   \author{A. Dolliou \inst{\ref{aff:mps}} 
    \and S. Mandal \inst{\ref{aff:mps}}
    \and K. Barczynski \inst{\ref{aff:davos}}
    \and T. Van Doorsselaere \inst{\ref{aff:leuven}}
    \and D. Berghmans \inst{\ref{aff:rob}}
    \and C. Froment \inst{\ref{aff:lpc2e}}
    \and F. Auch\`ere \inst{\ref{aff:ias}}
    \and P. Antolin \inst{\ref{aff:newcastle}}
    \and H. Eklund \inst{\ref{aff:esa}}
    \and Y. Zhu \inst{\ref{aff:eth}}
    \and E. Kraaikamp \inst{\ref{aff:rob}}
    \and C. Verbeeck \inst{\ref{aff:rob}}
      }

    \institute{%
        \label{aff:mps}{Max Planck Institute for Solar System Research, Justus-von-Liebig-Weg 3, 37077 Göttingen, Germany} \and \label{aff:davos}{Physikalisch-Meteorologisches Observatorium Davos, World Radiation Center, 7260 Davos Dorf, Switzerland} \and \label{aff:leuven}{Centre for mathematical Plasma Astrophysics, Mathematics Department, KU Leuven, Belgium} \and \label{aff:rob}{Solar-Terrestrial Centre of Excellence – SIDC, Royal Observatory of Belgium, Ringlaan -3- Av. Circulaire, 1180 Brussels, Belgium} \and \label{aff:lpc2e}{LPC2E, OSUC, Univ Orleans, CNRS, CNES, F-45071 Orleans, France}  \and \label{aff:ias}{Institut d’Astrophysique Spatiale, Bâtiment 121, Rue Jean Dominique Cassini, Université Paris Saclay, 91405 Orsay, France} \and \label{aff:newcastle}{School of Engineering, Physics, and Mathematics, Northumbria University, Newcastle upon Tyne, NE1 8ST, UK} \and \label{aff:esa}{European Space Agency (ESA), European Space Research and Technology Centre (ESTEC), Keplerlaan 1, 2201 AZ Noordwijk, The Netherlands} \and \label{aff:eth}{ETH-Zürich, Wolfgang-Pauli-Str. 27, 8093 Zürich, Switzerland}
    \\ \email{dolliou@mps.mpg.de} }

   \date{Accepted on 2026 July 15; published on}

  \abstract
   {Recent high-resolution observations from the Solar Orbiter mission can help detect the indirect signatures of heating at the smallest scales.} 
   {In this work we measure the properties and investigate the physical origin of propagating disturbances (PDs) in the intensity at the smallest resolvable scales with Solar Orbiter/EUI along coronal loops.}
   {We used two sequences of EUI/HRIEUV at high spatial (down to \SI{125}{\kilo\meter} per pixel) and temporal resolutions (\SI{5}{\second} of cadence). We placed slits along 13 active region (AR) coronal loops. We measured the plane-of-sky (PoS) velocities and the intensity perturbation damping along the slits of PDs. We also measured the periodicity of PDs in one slit by using a Fourier analysis.}
   {We report the detection of PDs that have high PoS velocities ranging between \SI{500}{} and \SI{2000}{\kilo\meter\per\second} (which we call "fast" PDs). They are only visible in the upper part of the coronal loops. The intensity increase associated with these fast PDs is on the order of 4\% to 8\% of the HRIEUV intensity, and we measured little to no damping of their intensity with distance. We also measured a peak in the Fourier power spectra at 2 min above the 95\% confidence limit that is associated with fast PDs. The fast PDs are detected in the same coronal loops as PDs with a lower PoS velocity (70 to \SI{90}{\kilo\meter\per\second}), which we refer to as "slow" PDs. Unlike the fast PDs, these slow PDs are only visible in the lower part of the coronal loop, and they show clear intensity damping. }
   {Slow PDs show properties consistent with slow magneto-acoustic modes or upflows. On the other hand, fast PDs cannot be explained by slow magneto-acoustic modes. Instead, they show properties consistent with fast flows induced by magnetic reconnection, current sheet generated by propagating transverse oscillations, and fast magnetohydrodynamics modes or \alfven{} waves.}

    \titlerunning{Fast propagating propagating disturbances along coronal loops in active regions}
    \authorrunning{Dolliou et al.}
   \maketitle
\nolinenumbers 
   
\section{Introduction}

Coronal loops in active regions (AR) can be heated to multiple million Kelvins (\SI{}{\mega\kelvin}s). Observations in the X-ray \citep[][]{Hannah_2008} and in the extreme ultraviolet ranges \citep[EUV;][]{Berghmans&Clette&Moses1998} combined with modeling \citep[][]{Ugarte_2019} suggest that the heating of the corona is impulsive (not continuous with time) and occurs through a large number of small-scale ($<$ \SI{e24}{\erg}) energy deposition events. The exact nature of the mechanisms responsible for energy transport and dissipation from the photosphere to the corona are still under discussion, but magnetic reconnection \citep[][]{Priest_2000} and waves \citep[][]{VanDoorsselaere_2020} likely have an important role.

 Intensity propagating disturbances (PDs) have been observed with the EUV for decades \citep{DeForest_Gurman_1998}. These PDs are of particular interest in the context of coronal heating because they might be the observational signature of processes responsible for energy (e.g., waves) or mass transfers (e.g., flows) from the photosphere to the corona. The PDs such as those detected by \cite{DeForest_Gurman_1998} along fan loops have a plane-of-sky (PoS) velocity close to the speed of sound in the corona \citep[][]{Kiddie_2012}. This property, along with the intensity damping with distance \citep[][]{Prasad_2014}, is consistent with slow-magneto-acoustic mode models \citep[][]{de_Moortel_2003}. Yet measurements with the EUV Imaging Spectrometer \citep[EIS;][]{Culhane_2007} on board the Hinode spacecraft \citep[][]{Kosugi_2007} have also shown blue wing enhancements consistent with periodic upflow \citep[][]{De_Pontieu_McIntosh_2010,Tian_2011} or associated with inherent slow-wave induced asymmetries \citep{verwichte2010}. Thus, the exact nature of the PDs remains unclear, especially given the similarities of the spectroscopic signatures between the slow wave and the periodic upflow models \citep[][]{De_Moortel_Antolin_Doorsselaere_2015}. 
 
 Despite this ambiguity, the wide range of PoS velocities reported for PDs in the past decades is still a good indicator that they might not share the same physical origin. Indeed, PoS velocities of PDs have been measured from the subsonic \citep[e.g., up to \SI{60}{\kilo\meter\per\second} seen in the EUV,][]{Mandal_2022} to sonic or supersonic ranges \citep[e.g., up to \SI{400}{\kilo\meter\per\second} seen in the Ly-$\alpha$ line,][]{Kubo_2016,Yoshida_2019} and all the way to Alfv\'enic values \citep[e.g., \SI{600}{\kilo\meter\per\second} seen in X-rays,][]{Cirtain_2007}. This indication is also supported by the fact that PDs with a wide range of PoS velocities can coexist in the same region \citep[see for instance,][]{Dolliou_2026,Zhang_2015,Cirtain_2007}. Furthermore, measurements of the periodicity of these PDs can also be used to investigate their drivers in the lower atmosphere or in the photosphere. As such, a three-to-five-minute period is often detected for PDs at the footpoints of AR loops \citep[][]{De_Moortel_2000}. Such periods have been interpreted as the result of p-mode leakage \citep[][]{De_Pontieu_2005}. Finally, observations of slow modes have also been used to derive plasma parameters, such as the adiabatic index and thermal conduction using coronal seismology techniques \citep[][]{vd2011}, or to locate the footpoints of magnetic loops \citep[][]{Rawat_2023} in order to derive the plasma-$\beta$ parameter \citep[][]{rawat2025}.

The objective of this work is to investigate PDs at the smallest resolvable scales in the EUV using observations from the Extreme Ultraviolet Imager \citep[EUI;][]{EUI_instrument} on board Solar Orbiter \citep[][]{Muller,Zouganelis2020} at high spatial and temporal resolutions. We report the detection with EUI of PDs with a wide range of PoS velocities simultaneously along the same AR coronal loops. We classified them into two categories depending on their PoS velocity range: "slow" PDs (\SI{80}{} to \SI{105}{\kilo\meter\per\second}) detected only near the coronal loop footpoints and "fast" PDs (\SI{500}{} to \SI{2200}{\kilo\meter\per\second}) detected in the upper part of the coronal loop. We also report the detection of fast PDs with the \SI{171}{\angstrom} channel of the Atmospheric Imaging Assembly \citep[AIA;][]{Lemen2012} on board the Solar Dynamics Observatory \citep[SDO;][]{Pesnell2012}.

\section{Observation}
\label{sec:obs}

\subsection{Extreme Ultraviolet Imager}
\begin{figure}
\centering
    \includegraphics{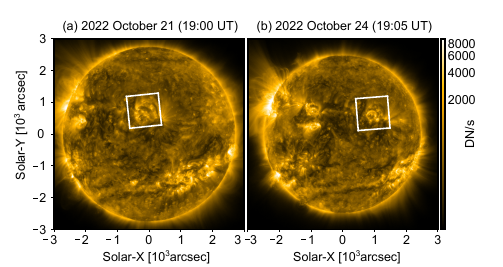}
    \caption{Context FSI 174 images for the (a) 2022 October 21 and (b) 24 sequences. The white rectangles show the field of view of HRIEUV at each date.}
    \label{fig:obs:context}
\end{figure}

We selected two sequences: one from 2022 October 21 (19:00 to 20:00 UT) and another from 2022 October 24 (19:05 to 19:30 UT; see Fig.\ref{fig:obs:context}). They are part of the Solar Orbiter Observing Plan \citep[SOOP;][]{Auchere_2020,Zouganelis2020} "R\_SMALL\_MRES\_MCAD\_AR-Long-Term", which was aimed at following the same AR over multiple days. The coordination included multiple instruments from Solar Orbiter, such as EUI. We used data from the \SI{174}{\angstrom} passband of the Full Sun imager (FSI 174) and from the High Resolution Imager EUV (HRIEUV). The passband of both imagers peaks with the \ion{Fe}{x} line (\SI{1}{\mega\kelvin}) but also includes contributions from transition region lines. During the two sequences, the field of view of HRIEUV included the AR tracked during the SOOP, as shown in Fig.\ref{fig:obs:context}. 

The pixel size of HRIEUV is equal to \SI{0.496}{\arcsec}, with a spatial resolution of two pixels \citep[][]{Berghmans_2023}. On 2022 October 21 and 24, the Solar Orbiter to Sun distances were equal to 0.35 and 0.39 AU, respectively. Thus, the pixel sizes were equivalent to \SI{125}{\kilo\meter} and \SI{140}{\kilo\meter} on the solar atmosphere, given a reprojection on a sphere of radius 1.004$R_\mathrm{sun}$. This radius corresponds to a height of $\approx$\SI{3}{\mega\meter} above the photosphere, where the average temperature of the atmosphere \citep[$\approx$ \SI{1}{\mega\kelvin},][]{Aschwanden2005} reaches the temperature response peak of HRIEUV. The HRIEUV cadence was set to \SI{5}{\second} for the two sequences.

We used Level-2 (L2) FITS files from the EUI data release 6.0 \citep[][]{euidatarelease6} for the two sequences from 2022 October 21 (19:00 to 20:00 UT) and 24 (19:05 to 19:30 UT). The pointing information in the metadata of the FSI 174 FITS files have been corrected previously with a limb fitting method during the Level-1 (L1) to L2 processing. We corrected the pointing information on the HRIEUV FITS files by co-aligning them with the FSI 174 files that are closest in time. To do so, we used the co-alignment Python package  \software{euispice\_coreg} described in Appendix A of \cite{Dolliou_2024}. We also applied a jitter correction to the HRIEUV sequence using a method inspired from Appendix A of \citet{Chitta_2022}: The HRIEUV sequence was divided into \SI{1}{\min} sublists overlapping by one image. Each image was co-aligned (in Carrington coordinates) with the first image of its sublist using \software{euispice\_coreg}, resulting in a correction of the jitter over the whole sequence. Finally, each HRIEUV image was reprojected into the frame of the first image of its sequence, removing the effect of the Carrington rotation. This step was performed with the \software{reproject} and the \software{sunpy} Python packages \citep{sunpy_community2020}.   

\subsection{Atmospheric Imaging Assembly}

For this work, we chose to use the AIA sequence on 2022 October 24, as the location of the tracked AR had a lower viewing angle from Earth’s perspective on October 24 ($\approx$\SI{-20}{\degree}) compared to October 21 ($\approx$\SI{-60}{\degree}). On the October 24, the time delay between HRIEUV and AIA 171 images due to the difference in light travel time was \SI{520}{\second}. We used a dataset (19:05 to 19:45 UT) of the \SI{171}{\angstrom} channel of AIA, which has a temperature response peaking at \SI{0.9}{\mega\kelvin}. This temperature sensitivity peak is close to the one of HRIEUV at \SI{1}{\mega\kelvin}. The AIA 171 dataset was prepared with the \software{aiapy} open project \citep[][]{Barnes2020}. The Carrington and the differential rotations were removed by reprojecting each image onto the frame of the image at 19:05 UT. The reprojection was done with the \software{reproject} and \software{sunpy} Python packages.

\section{Results}
\label{sec:results}

\subsection{Plane-of-sky velocity measurements}
\label{sec:results:pos_velocity}

\begin{figure*}
    \centering
    \includegraphics{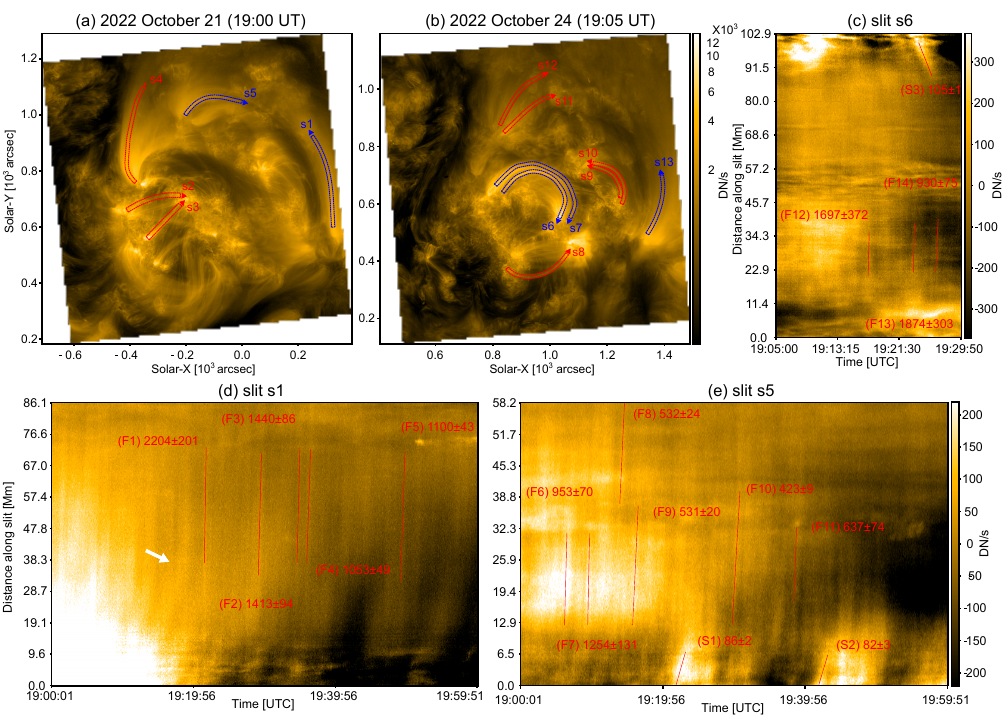}
    \caption{Detection of the fast and slow PDs along coronal loops. HRIEUV images from the (a) 2022 October 21 and (b) 24 sequences showing the locations of the 13 slits. The slit colors indicate whether fast PDs were detected (blue) or not (red) on the time distance maps. The arrows show the direction taken as convention for the increasing distance. Time distance maps along the slits are shown for (c) s$_6$, (d) s$_1$, and (e) s$_5$. Each row of the time distance map has been subtracted by its temporal mean over the whole sequence. The PoS velocities of fast and slow PDs are given in kilometer per second, and the associated slope is shown as a red line. The labels of the fast and slow PDs are provided within parentheses. The white arrow in (d) highlights an example where slow and fast PDs appear to connect.}
    \label{fig:results:fov_time_distance_maps_rundiff_vel}
\end{figure*}

\begin{figure}
    \centering
    \includegraphics{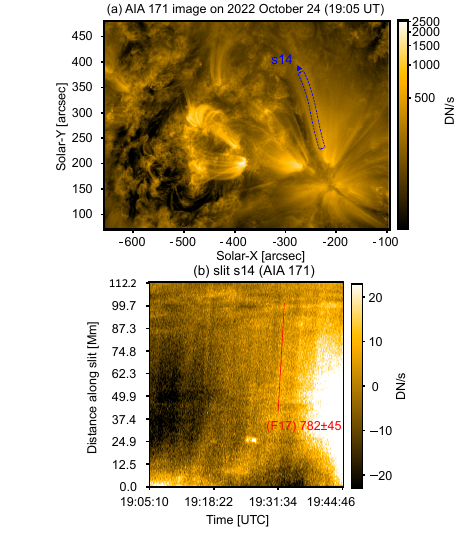}
    \caption{Detection of fast PDs along coronal loops with AIA 171. (a) Image of AIA 171 on 2022 October 24 at 19:05 UT showing a slit labeled s$_{14}$ with a width of \SI{12}{\arcsec} placed at the approximate location of s$_{13}$ as seen from the Solar Orbiter viewpoint. (b) Time distance map computed along s$_{14}$, where each row has been subtracted by its temporal mean over the whole sequence. The fast PD labeled F$_{17}$ with a PoS velocity of \SI[separate-uncertainty = true]{782(45)}{\kilo\meter\per\second} is shown in red. The PoS velocity has been computed in a similar way as for the fast PDs shown in Fig.\ref{fig:results:fov_time_distance_maps_rundiff_vel}.}
    \label{fig:results:aia171_fastpds_v3}
\end{figure}

We placed multiple slits along AR loops, labeling them as s$_1$ to s$_{13}$ (Fig.~\ref{fig:results:fov_time_distance_maps_rundiff_vel}a, b). The color of the slits indicates whether fast PDs were clearly detected (blue) or not (red) in HRIEUV. The slits were placed along strands identified in images enhanced by the \software{wavelet-optimized whitening} algorithm \citep[WOW,][]{Auchere_2023}. We computed the time-distance maps along the slits (Fig.~\ref{fig:results:fov_time_distance_maps_rundiff_vel}c, to e, Fig.~\ref{fig:annex:AR21_AR24_all_other_slits_time_distance_v4}). For visualization purposes, we subtracted from each row of the time distance map the temporal mean over the whole sequence. We computed the PoS velocity of 20 PDs with a well-defined peak above the background using a method described in Appendix A of \cite{Dolliou_2026}. This method can be summarized as follows: Equidistant locations separated by \SI{5}{pixels} were selected on the time distance maps along the slit. A Gaussian fit was performed at each location to compute the central times of the intensity peak. A linear fit between the central times and the distances along the slit was then used to estimate the PoS velocity and its uncertainty. We note that all PDs studied in this work have a PoS velocity that is temporally resolved by HRIEUV at a \SI{5}{\second} cadence. Fast PDs are only detected in the upper part of the coronal loop bundles, while slow PDs are detected near the loop footpoints. We note that the 20 PDs studied in this work are not the only PDs detected during these two sequences. 

The fast PDs were mainly detected in slits s$_1$, s$_5$, s$_6$ (Fig.~\ref{fig:results:fov_time_distance_maps_rundiff_vel}), and s$_{13}$ (Fig.\ref{fig:annex:AR21_AR24_all_other_slits_time_distance_v4}) during the entirety of the two sequences. They were not clearly detected in every coronal loop bundle we studied though, such as the ones covered by s$_3$ and s$_{8}$ (Fig.~\ref{fig:annex:AR21_AR24_all_other_slits_time_distance_v4}). We measured the PoS velocities for 16 of them, which we labeled from F$_1$ to F$_{16}$. The fast PDs are associated with a low intensity increase compared to the background. Their PoS velocities range from \SI{500}{} to \SI{2200}{\kilo\meter\per\second} (Table \ref{table:events}). The fast PDs studied in this work appear to mostly propagate in the same direction, from one of the footpoints to the upper part of the coronal loop. Finally, we checked that the fast PDs are not artifacts produced by the background intensity or the strands moving in and out of the slits, which have a limited width of \SI{12}{\arcsec} (Appendix \ref{sec:annex:verify_artifacts_transverse}). We also verified that the fast PDs are not the result of instrumental artifacts (Appendix \ref{sec:annex:verify_artifacts_instrumental:detection_fastpd}).

Unlike the fast PDs, the slow PDs were only detected near the footpoint of the coronal loop. For instance, they are only detected below \SI{20}{\mega\meter} in s$_1$ and in s$_5$ (Fig.~\ref{fig:results:fov_time_distance_maps_rundiff_vel}d, e). They also propagate toward the upper part of the coronal loop. The slow PDs were well detected in s$_1$, s$_2$, s$_4$, s$_5$, s$_6$, s$_7$, and s$_{13}$ (Fig.~\ref{fig:results:fov_time_distance_maps_rundiff_vel}c to e and Fig.~\ref{fig:annex:AR21_AR24_all_other_slits_time_distance_v4}). We note that not all slow PDs are highlighted by red lines in the figures, as we measured the PoS velocity for four of them only, which we named S$_1$ to S$_4$. The values range between \SI{82}{} and \SI{105}{\kilo\meter\per\second} and are provided in the time distance maps and in Table \ref{table:events}. We also noticed examples where slow and fast PDs appear to connect. For instance, the slow PD highlighted by a white arrow in Fig.~\ref{fig:results:fov_time_distance_maps_rundiff_vel}d appears to show an acceleration and to end up as a fast PD. However, due to the ubiquity of both fast and slow PDs during the entirety of the time sequences, these connections could be caused by chance. We also note that the increase of the apparent speed of slow PDs could be induced by the curvature of the coronal loop and projection effects \citep[][]{Sieyra_2022}.    

We note that fast PDs are difficult to identify directly in base-difference HRIEUV movies, likely due to their low intensity increase above the background compared to the noise. The movie "F1\_base\_difference.mp4" shows as an example the identification of F$_1$ along slit s$_1$ (in blue) in a base-difference movie with the reference time taken at 19:20:51 UT. The associated movie is available online. The time-distance map of s$_1$ zoomed-in around F$_1$ is also shown, with each time step indicated by a vertical dotted blue line. In the movie, there is an intensity increase at 19:21:26 UT along s$_1$ at \SI{760}{\arcsec} in solar-Y (corresponding to a distance of \SI{40.8}{\mega\meter} along the slit in Fig.\ref{fig:results:fov_time_distance_maps_rundiff_vel}d). This is followed by an intensity increase all along the slit up to 19:21:46 UT, which is consistent with the time appearance of F$_1$ at 19:21:26 UT and with its lifetime of few time steps in the time distance map. In addition, we estimated the coherence width of F$_1$ by performing three perpendicular cuts across s$_1$ at 19:21:26 UT on the latitudes (solar-Y) of \SI{760}{\arcsec}, \SI{765}{\arcsec}, and \SI{770}{\arcsec}. We estimated a full width at half maximum associated with the fast PD to be on the order of 1 to \SI{2}{\mega\meter}. 

Finally, we also report the detection of fast PDs with AIA 171. To do so, we placed the slit called s$_{14}$ on AIA 171 images along a strand that appeared close to the location where s$_{13}$ is placed on the HRIEUV image (Fig.\ref{fig:results:aia171_fastpds_v3}a). We estimated the PoS velocity of one fast PD (labeled F$_{17}$) on the \SI{12}{\arcsec} width slit to be equal to \SI[separate-uncertainty = true]{782(45)}{\kilo\meter\per\second} (Fig.\ref{fig:results:aia171_fastpds_v3}b).  Given the difference of viewpoints between Solar Orbiter and SDO, this value is reasonably close to the ones of F$_{15}$ at \SI[separate-uncertainty = true]{978(72)}{\kilo\meter\per\second} and F$_{16}$ at \SI[separate-uncertainty = true]{695(45)}{\kilo\meter\per\second} measured with HRIEUV on s$_{13}$ (Fig.~\ref{fig:annex:AR21_AR24_all_other_slits_time_distance_v4}). Thus, it is likely that we detected similar features with HRIEUV and AIA 171 in that case. We verified that these fast PDs detected with AIA 171 are not caused by transverse oscillations of strands going in and out of the slit in Appendix \ref{sec:annex:verify_artifacts_transverse}.  

\subsection{Damping analysis}
\label{sec:results:damping}

\begin{figure}
    \centering
    \includegraphics{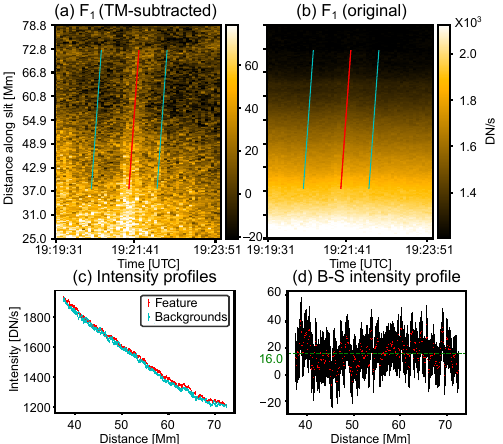}
    \caption{Damping analysis of F$_1$ detected on $s_1$ (Fig.~\ref{fig:results:fov_time_distance_maps_rundiff_vel}d). The time distance map along $s_1$ is zoomed-in around F$_1$. In (a), each row of the time distance map has been subtracted by the temporal mean (TM) over the whole sequence, while in (b) it has not. The red line is the same as the one shown in Fig.~\ref{fig:results:fov_time_distance_maps_rundiff_vel}d. The cyan lines were used to estimate the intensity profiles of the background. The intensity profiles of F$_1$ and of the backgrounds (c) have been computed along the red and cyan lines, respectively. The background subtracted (B-S) intensity profile of F$_1$ is shown in (d). The values are shown as red points and the uncertainties as black bars. The average of the intensity profile (\SI{16.0}{\dn\per\second}) over the whole distance is displayed as a green dotted horizontal line.}
    \label{fig:results:AR_21_10_22T1900_s1_damping_0}
\end{figure}

In this section, we measure how the intensity profiles of fast and slow PDs vary with distance along the slit compared to the background intensity. We performed this damping analysis on nine fast PDs (F$_1$ to F$_8$ and F$_{14}$) and one slow PD (S$_1$). We illustrate the method we used by applying it to F$_1$. First, we computed the intensity profile of F$_1$ along the red line obtained from the PoS velocity computation (Fig.~\ref{fig:results:AR_21_10_22T1900_s1_damping_0}a, b). Two background intensity profiles were computed along lines parallel to the red lines and temporally shifted before and after the F$_1$ in the time distance map (Fig.~\ref{fig:results:AR_21_10_22T1900_s1_damping_0}a, b). The intensity profiles of F$_1$ and of the two backgrounds are given in Fig.~\ref{fig:results:AR_21_10_22T1900_s1_damping_0}c. A linear interpolation between the two background profiles was used to estimate the one at the time of F$_1$. The resulting background-subtracted intensity profile of F$_1$ is shown in Fig.~\ref{fig:results:AR_21_10_22T1900_s1_damping_0}d.

The background-subtracted intensity profile of F$_1$ does not stray away more than its uncertainty from its average value of \SI{16}{\dn\per\second} over the \SI{30}{\mega\meter} distance. This is also true for all other fast PDs detected (Fig.~\ref{fig:annex:damping_all_others_s1} and Fig.~\ref{fig:annex:damping_s5_s6}). Therefore, we did not detect any evidence of intensity variation over the distance where the fast PD was measured. The only exception is F$_8$ (Fig.~\ref{fig:annex:damping_s5_s6}f), where we measured an intensity decrease from 40 to \SI{10}{\dn\per\second} over \SI{18}{\mega\meter}. On the other hand, the background-subtracted intensity profile of the slow PD s$_1$ shows a clear decrease in value from \SI{250}{\dn\per\second} in the footpoint (at $D=$\SI{98}{\mega\meter}) to \SI{50}{\dn\per\second} higher up in the coronal loop ($D=$\SI{89}{\mega\meter}, Fig.~\ref{fig:results:AR_21_10_22T1900_s2_damping_2}).

Finally, we also estimated the percentage of intensity increase associated with the fast PDs, which was useful for our later comparison with wave models (Section \ref{sec:discussion}). However, we wished to mitigate the impact of the filling factor when averaging over the \SI{12}{\arcsec} widths of the slits. To do so, we performed a damping analysis similar to the one described in this section but over sub-slits of 1 pixel width only (compared to 24 pixels used previously; see the method described in Appendix  \ref{sec:annex:verify_artifacts_instrumental:detection_fastpd}). We estimate the intensity increases associated with fast PDs to be on the order of 2 to 4\% of the total intensity. For instance, we measured an average of \SI{118}{\dn\per\second} for the background-subtracted intensity profile of F$_{14}$, with a total intensity ranging between \SI{4500}{} and \SI{3000}{\dn\per\second} (see Fig.~\ref{fig:annex:check_rounding_s6}). Also, one should estimate the contribution of the foreground and background along the line of sight (LOS) to the total intensity. For slit s$_6$, we estimate it to be equal to about $\approx50\%$ of the total intensity (Fig.\ref{fig:annex:time_distance_bckg}). We thus estimate the intensity fluctuations associated with fast PDs to be on the order of 4\% to 8\% of the total intensity of the strand. 

\subsection{Periodicity analysis of fast PDs}
\label{sec:results:periodicity}

\begin{figure}
    \centering
    \includegraphics{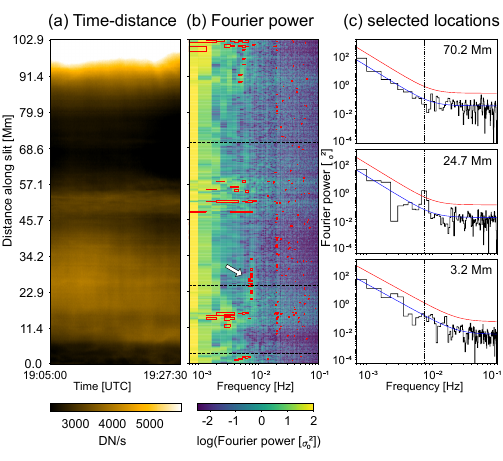}
    \caption{Periodicity analysis along slit s$_6$. (a) Time-distance map computed along s$_6$. Here, the map is shown without subtraction of the average at each column. (b) Fourier power map obtained by computing the Fourier power for each row of the time distance map. The red contours indicate the pixels above the 95\% confidence level. The white arrow highlights the Fourier peaks of interest discussed in Sect. \ref{sec:results:periodicity}. Fourier power for selected rows along the time distance map are shown in (c). The blue and red lines are respectively the Fourier power of the background fitted from a power law and the 95\% confidence level. The black vertical dotted line indicates the \SI{130}{\second} period ($7\cdot10^{-3}$ Hz).  }
    \label{fig:results:AR24_s6_periodicities}
\end{figure}

We verified if the fast PDs are detected on a periodic pattern. For s$_6$, we computed the Fourier power at every row of the time distance map (Fig.~\ref{fig:results:AR24_s6_periodicities}a,b). The method we used to compute the Fourier power and the 95\% confidence levels are detailed in Appendix \ref{sec:annex:FFT}. We noticed peaks near the $7\cdot10^{-3}$ Hz frequency ($\approx$ \SI{2}{\min} period) between the \SI{20}{\mega\meter} and \SI{35}{\mega\meter} locations along the slit (highlighted by a white arrow in Fig.~\ref{fig:results:AR24_s6_periodicities}b). Many of the peaks are above the 95\% confidence interval (see, e.g., the middle panel of Fig.~\ref{fig:results:AR24_s6_periodicities}c). This length interval of [20, 35] \SI{}{\mega\meter} corresponds to where fast PDs are best detected in HRIEUV along s$_6$ (Fig.~\ref{fig:results:fov_time_distance_maps_rundiff_vel}c and \ref{fig:results:AR24_s6_periodicities}a). We noticed that no other significant peaks in the Fourier power map could be detected with such a spatial coherence of more than \SI{5}{\mega\meter} along the slit (those at $2\cdot10^{-2}$ Hz are caused by instrumental artifacts; see Appendix \ref{sec:annex:verify_artifacts_instrumental:detection_period}). This is why these peaks at \SI{2}{\min} are likely to be associated with the fast PDs and not to some other unrelated emission of the coronal loop bundle. 

We also performed a similar periodicity analysis on the slits s$_1$ and s$_5$. The fast PDs appear regularly (Fig.\ref{fig:results:fov_time_distance_maps_rundiff_vel}d, e), but we did not detect peaks above the 95\% confidence level that could be attributed to fast PDs. As such, the period of \SI{2}{\min} we found for s$_6$ does not appear to be a result for all slits. We note that slit s$_1$ and s$_5$ cover fan loops, while s$_6$ covers a smaller coronal loop bundle. Stronger LOS integration effects for fan loops might explain the lack of a clear peak in the Fourier spectrum in s$_1$ and s$_5$.

\section{Discussion}
\label{sec:discussion}

In this work, we have reported on the detection of slow and fast PDs along coronal loops in AR with HRIEUV. Slow PDs show properties typical of those first reported in \cite{DeForest_Gurman_1998} in terms of PoS velocity close to the sound speed \citep[$c_\mathrm{s} = $ \SI{152}{\kilo\meter\per\second} at \SI{1}{\mega\kelvin},][]{Priest_1982} and of intensity damping with distance \citep[][]{Prasad_2014}. These properties can be reproduced with a model of slow magneto-acoustic mode \citep[][]{de_Moortel_2003}. The fast PDs we studied in this work, on the other hand, cannot be explained by slow modes. They show a PoS velocity ranging between \SI{500}{} and \SI{2200}{\kilo\meter\per\second} (Table \ref{table:events}) and no detectable intensity damping with distance (with the exception of F$_8$; see Fig.\ref{fig:annex:damping_s5_s6}e). Their intensity fluctuation is on the order of 4\% to 8\% of the HRIEUV intensity. Also, the fast PDs we studied mainly appear to propagate in the same direction, from the coronal loop footpoint toward its upper part. In one case out of the three slits we analyzed, we could associate them with a peak at \SI{2}{\min} in the Fourier power spectra. 

Similar instances of coexisting slow and fast PDs have been reported before in AR loops during flares with EUV imagers \citep[][]{Zhang_2015} or under different magnetic configurations, such as along network loops \citep[][]{Dolliou_2026}. However, to our knowledge, this is the first time such ubiquitous, periodic, and fast PDs are detected with an EUV imager along nonflaring AR coronal loops. We also want to highlight that the fast PDs are not only detected in HRIEUV but also in AIA 171 (Section \ref{sec:results:pos_velocity}). Despite the lower signature, the detection of fast PDs in AIA 171 is nevertheless important, as it can lead to follow-up statistical studies taking advantage of the large database accumulated by AIA over the past decade.

In the following, we discuss the nature of the fast PDs seen in HRIEUV in terms of plasma flows, pure thermal effect (Section \ref{sec:discussion:plasma_flows_thermal}), magnetohydrodynamics (MHD) waves (Section \ref{sec:discussion:mhd_waves}) and coronal loop geometrical modulations (Section \ref{sec:discussion:loop_structural_modulation}). These processes are nonexclusive and can be intertwined, but we still consider them one by one for the sake of the discussion. We also focus on the possible drivers of these fast and slow PDs at the coronal loop footpoints (Section \ref{sec:discussion:driver_footpoint}). 

\subsection{Plasma flows and thermal effects}
\label{sec:discussion:plasma_flows_thermal}

Average plasma flows in AR are measured with Doppler shift measurements of coronal \citep[$\sim$ \SI{30}{\kilo\meter\per\second},][]{Zhu_2025} and flaring lines \citep[up to $\approx$ \SI{300}{\kilo\meter\per\second},][]{Antonucci_1985}. We note however that intensity peak motions attributed to plasma flows can reach PoS velocities close to those of fast PDs \citep[up to \SI{600}{\kilo\meter\per\second},][]{Cirtain_2007}. Indeed, plasma flows can be accelerated to values close to the Alfv\'en speed during a reconnection event. This model could explain fast PDs if periodic reconnection was happening near the footpoint of the coronal loop (see Sect. \ref{sec:discussion:driver_footpoint}).

With regard to thermal effects, we note that fast PDs are separated by time scales on the order of 2 min. However, a single coronal loop in AR would not have the time to cool down and heat up in such a short time frame. Indeed, conduction and radiation time scales for typical AR coronal loops are on the order of $\approx$ \SI{30}{\min} to a few hours \citep[$L=$ \SI{e10}{\centi\meter}, $n = $ \SI{e10}{\per\centi\meter\tothe{3}}, $T=$ \SI{3e6}{\kelvin},][]{Cargill_2004}. Therefore, assuming fast PDs are caused by thermal effects, we suggest each individual fast PD to be associated with a single distinct magnetic loop along the LOS. Scenarios that could explain the initial heating of the magnetic loop include: thermal conduction \citep[see for instance the fast temperature increase in Fig.2b of][]{Fang_2015} or the generation of current sheets by transverse oscillations propagating at \alfven{} speed  \citep[][]{de_pontieu_2017}. While originally applied to spicules, the effect described by \cite{de_pontieu_2017} is not limited to the lower atmosphere. The current sheets can also reach the corona and heat it, as recently shown by \cite{Chen_2026} during the simulation of a blowout jet.

\subsection{MHD waves}
\label{sec:discussion:mhd_waves}

\begin{figure*}
\sidecaption
\includegraphics[width=12cm]{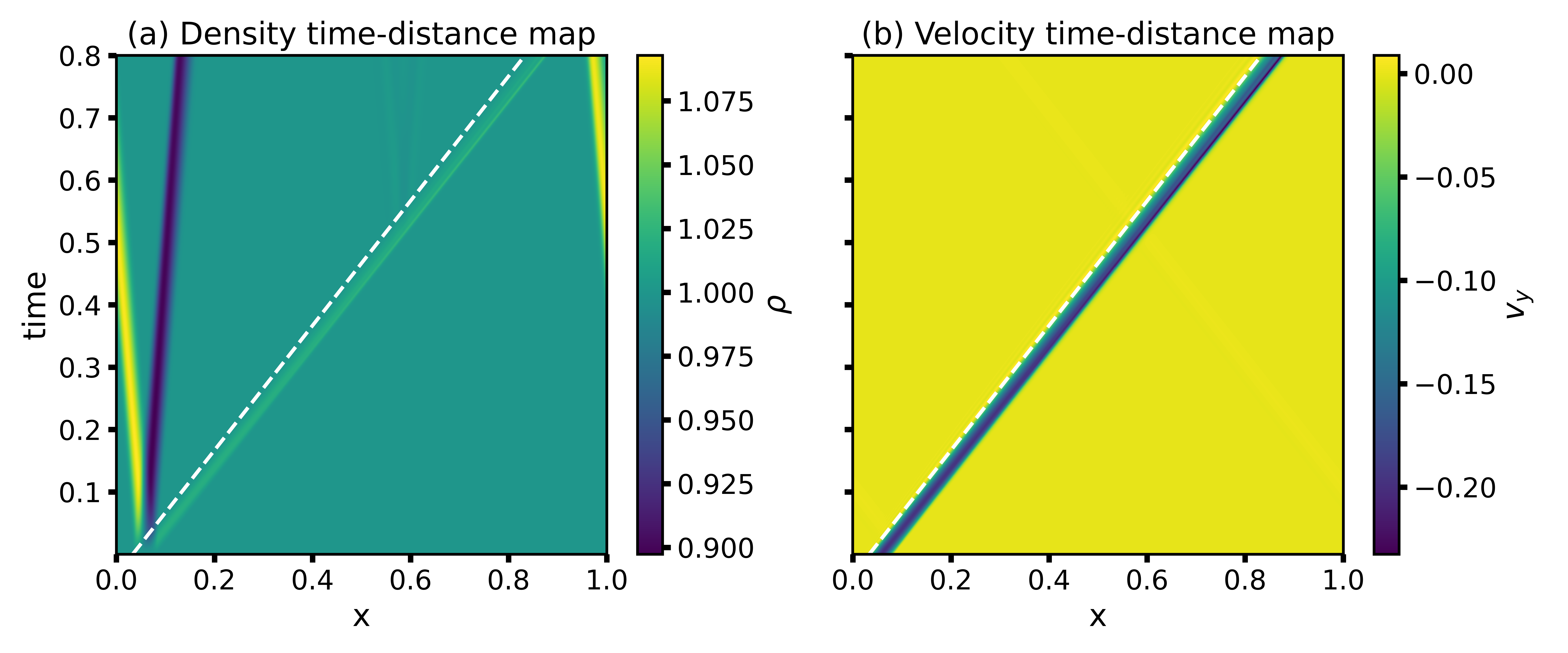}
	\caption{Propagation of a density perturbation driven by the ponderomotive force in an \alfven{} wave. The left panel shows the time-distance map  along the $x$-axis of the density $\rho$, while the right panel shows the graph for the transverse velocity displacements $v_y$. The latter have been normalized to the background \alfven{} speed. The white dashed line shows a line propagating with the \alfven{} speed in the initial setup. For more details on the nomenclature and on the simulation setup, see Appendix \ref{sec:annex:ponderomotive_force:python}.  }
	\label{fig:pythonsims}
\end{figure*}

Here we consider the possibility that fast PDs could be explained by MHD waves. The PoS velocities of fast PDs are similar to the Alfv\'en speed in the corona \citep[$\sim$\SI{1000}{\kilo\meter\per\second},][]{Tomczyk_2007}. As such, they could be the signature of fast MHD modes, which are compressive waves with a phase speed higher than the \alfven{} speed. Even without considering nonlinear effects (e.g., Kelvin Helmholtz instability), fast MHD modes can induce perturbation in the coronal loop density of a few percent \citep[][]{Antolin_2017,Mondal_2024}. We note however that no large-scale transverse oscillation could be measured in our case (Fig.\ref{fig:annex:checking_fast_waves_s1}, \ref{fig:annex:checking_fast_waves_s6}, h to j). This is inconsistent with the prediction from the forward simulation results by \cite{Antolin_2017}, unless the transverse oscillation is polarized along the LOS.

To compare these perturbations to our observational results, we assumed the intensity to scale with the density squared $I\sim n^2$. The intensity fluctuations $\frac{\delta I}{I}$ relate to those of the density $\frac{\delta \rho}{\rho}$ as follows (see also Appendix \ref{sec:annex:ponderomotive_force:analytical}):
\begin{equation}\label{eq:di_drho}
	\frac{\delta I}{I} \approx 2 \frac{\delta \rho}{\rho}.
\end{equation}
As such, the intensity fluctuations of a few percent that we measured are consistent with density fluctuations induced by fast MHD modes. In particular, sausage modes are possible candidates to explain the fast PDs. Indeed, they can also be associated with intensity increase on the order of a few percent to 20\%, depending on the viewing angle and on the spectral line \citep[][]{Antolin_2013}. However, fast sausage mode pulses are expected to show strong dispersion and wave leakage \citep[][]{Shi_2026}. Thus, the sausage mode model appears to not be consistent with the little to no damping we measured for fast PDs along their propagation path (Section \ref{sec:results:damping}). We note that the width fluctuations of the strand are expected to be low for a sausage mode and probably below the instrumental resolution \citep[$\approx$5\% of the half width in][]{Antolin_2013}. This is why it is not surprising that we did not detect any fluctuation in the apparent width of the strands as seen in HRIEUV (Fig.\ref{fig:annex:checking_fast_waves_s1}, \ref{fig:annex:checking_fast_waves_s6}, h to j). Nonetheless, the apparent lack of transverse oscillation is a good argument for rejecting the hypothesis that fast PDs are caused by kink waves.

Torsional \alfven{} waves have been observed in large numbers in coronal AR loops \citep[][]{Tomczyk_2007,Morton2026}. They are incompressible by nature and are unlikely to be detected by an EUV imager. However, wave modifications due to nonuniformity can lead to potential signatures in imagers (e.g., wave steepening into shock, nonlinear phase mixing, or mode conversions into sausage or slow MHD modes). In particular, fast PDs could be the signature of a density front propagating at the \alfven{} speed through the ponderomotive force \citep{Shestov_2017,Touresse_2024}. Such a density front can appear as a fast PD when observed with HRIEUV. However, given their respective numerical setups, previous works employing simulations have indicated very low density fluctuations above the background associated with this effect ($\delta\rho/\rho\sim 0.0001$ in \citet{Shestov_2017}). The question remains whether such an effect could explain the 4\% to 8\% intensity increase in HRIEUV that we measured for fast PDs. This is why to test this theory we performed analytical derivations and 1.5D numerical simulations of a propagating \alfven{} wave (see Appendix \ref{sec:annex:ponderomotive_force}). The results showed that the intensity fluctuations (4\% to 8\%) and the PoS velocities (\SI{500}{} to \SI{2200}{\kilo\meter\per\second}) we measured for fast PDs are consistent with a density front driven by an \alfven{} wave of transverse velocity amplitudes ranging between \SI{440}{} and \SI{620}{\kilo\meter\per\second}. The results from the 1.5D MHD simulation confirm that a 2\% density fluctuation can propagate with \alfven{} speed (Fig.\ref{fig:pythonsims}a) along with an \alfven{} wave of \SI{440}{\kilo\meter\per\second} transverse velocity amplitude (Fig.\ref{fig:pythonsims}b). For comparison, the transverse velocity amplitudes of the ubiquitous small-scale torsional \alfven{} waves detected in coronal AR loops have been measured to be up to \SI{10}{\kilo\meter\per\second} \cite[][]{Tomczyk_2007,Morton2026}. Also, larger scale torsional \alfven{} waves have been measured to have an amplitude of up to \SI{60}{\kilo\meter\per\second} \citep[][]{Petrova_2024}. However, numerical works have shown that such amplitude measurements are strongly impacted by LOS integration effects \citep[by at least by a factor 10,][]{Pant_2019}. Further, assuming the  torsional \alfven{} waves originate from the chromosphere, a \SI{400}{} to \SI{600}{\kilo\meter\per\second} transverse velocity amplitude in the corona would only imply a \SI{20}{} to \SI{30}{\kilo\meter\per\second} amplitude in the chromosphere. Here, we assumed energy conservation with a density and a magnetic field decrease by a factor of \SI{e3}{} and \SI{10}{} from the chromosphere to the corona, respectively. We conclude that torsional \alfven{} waves with a propagating density front are a promising lead to explain fast PDs. 

Finally, the little to no damping we measured for most fast PDs indicates low energy dissipation in the corona through thermal conduction or compressive viscosity. The mechanism behind fast PDs is unlikely to significantly heat the corona at the height where we detected them with HRIEUV. However, by assuming fast PDs to be the signature of fast MHD modes, we can still compare their kinetic energy input to the one necessary to heat the corona (assuming they become damped at a greater height) or to drive the solar wind. We used the following formula to estimate the kinetic energy flux $E_\mathrm{kin}$ \citep{vd2014}:   
\begin{equation}\label{eq:ekin}
    E_\mathrm{kin} \approx f \rho \delta v_\mathrm{tr}^2 v_\mathrm{PoS}.
\end{equation}
We set the density to typical values in ARs ($\rho=$~\SI{e-15}{\gramme\per\centi\meter\tothe{3}}). We set the filling factor $f$ to one, making our estimation of the kinetic energy flux an upper limit. We took the boundary values for the PoS velocities ($v_\mathrm{PoS}$; \SI{500}{} to \SI{2200}{\kilo\meter\per\second}) and for the intensity fluctuations (4\% to 8\%). For the fast MHD mode model, the transverse velocity fluctuations ($\delta v_\mathrm{tr}$) can be estimated using Eq. \ref{eq:di_drho} and assuming $\delta \rho/\rho \sim \delta v/v$ \citep[Eq. 30 in][]{Shestov_2017}. In that case, the kinetic energy flux is estimated to range between $E_\mathrm{kin}\approx$ \SI{5e4}{} and \SI{2e7}{\erg\per\centi\meter\tothe{2}\per\second}. These values are barely enough to compensate for the coronal energy losses in ARs \citep[$\sim$ \SI{e7}{\erg\per\centi\meter\tothe{2}\per\second},][]{Withbroe_1977}. However, the kinetic energy input is within the range of the energy fluxes required to sustain the fast \citep[$\sim$ \SI{e5}{\erg\per\centi\meter\tothe{2}\per\second},][]{Bale_2023} and the slow solar winds \citep[$\sim$ \SI{e5}{} to \SI{e6}{\erg\per\centi\meter\tothe{2}\per\second},][]{Wang_1994}. This encourages future follow-up studies to evaluate if fast PDs can be detected on open field regions. In particular, small-scale solar wind features with high PoS velocities (up to \SI{520}{\kilo\meter\per\second}) have been detected recently in the middle corona by \citet{Zhukov_2026} with the Association of Spacecraft for Polarimetric and Imaging Investigation of the Corona of the Sun \citep[ASPIICS;][]{Zhukov_2025} on board the Proba-3 spacecraft. An interesting perspective would be to imply that these fast PDs could be the lower coronal equivalent of the small-scale solar wind dynamics.

\subsection{Coronal loop geometrical modulation}
\label{sec:discussion:loop_structural_modulation}

 In addition to the interpretations discussed above, we note that geometrical modulations within the coronal loop bundle may also play a role in the observed fast PD signatures. In particular, small changes in the coronal loop geometry can modify the LOS integration depth through the emitting plasma and therefore the observed intensity. This could produce intensity features propagating with apparent high phase speeds. Similarly, misalignment between the slit and the coronal loop axis, combined with transverse motions across the PoS could generate artificial propagating intensity disturbances in the time-distance maps. These are geometrical changes and are independent from any change in the coronal loop temperature, density, and emissivity. These modulations of the coronal loop geometry can have a wide variety of drivers (e.g., magnetic reconfiguration, perturbations). In particular, MHD waves can induce changes in the LoS integration depth through the emitting plasma \citep[see][for kink and sausage modes]{Cooper_2003}. This could produce fast propagating intensity peak motions close to Alfv\'enic speed, such as those seen in HRIEUV.

\subsection{Origin of fast PDs at the coronal loop footpoints}
\label{sec:discussion:driver_footpoint}

Hypotheses about the origin of fast PDs open up new perspectives.
Fast PDs are mainly visible in the upper part of the coronal loops. However, as they appear to propagate in the same direction, from one footpoint to the top of the loop, we suggest that their origin is likely to be located in the lower part of the loop (lower corona, transition region, or chromosphere). In this picture, fast PDs could not be clearly identified in the lower part of the coronal loop because of the strong background intensity fluctuations at these low heights (in particular due to slow PDs). This hypothesis is strengthened by observational evidence, such as the fast PDs F$_{10}$ and F$_{11}$ in slit s$_5$ (Fig.\ref{fig:results:fov_time_distance_maps_rundiff_vel}d), which appear to be associated with intensity fluctuations in the lower part of the coronal loop down to \SI{1}{\mega\meter} and \SI{6.5}{\mega\meter}, respectively.

Magnetic reconnection in the lower part of the coronal loops could generate jets \citep[][]{Pariat_2010,Wyper_2016,nobrega_sivero_2025}, fast MHD modes \citep[][]{Mondal_2024}, torsional \alfven{} waves \citep[][]{Touresse_2024}, thermal conduction, or current sheets induced by magnetic perturbation (e.g., \alfven{} waves) propagating at \alfven{} speed  \citep[][]{de_pontieu_2017}. The period of \SI{2}{\min} we measured could be related to p-mode-driven \citep[\SI{5}{\min},][]{Leighton_1962} magneto-acoustic shocks in the chromosphere, which could induce reconnection or the generation of magnetic perturbations in the lower part of the loop. The period of \SI{2}{\min} could also be linked to that of granulation \citep[\SI{5}{} to \SI{10}{\min},][]{Hirzberger_1999}. Indeed, regular flux emergence could induce periodic reconnection at the coronal loop footpoints involving small-scale magnetic structures \citep[][]{Chitta_2017}. From this perspective, fast PDs could be the signature of the energy releases in the lower part of the coronal loops. Another possible mechanism involved in the generation of fast PDs could be mode conversion, such as the linear slow-fast mode conversion in chromosphere at the equipartition layer, where the \alfven{} and the sound speeds coincide \citep[][]{Cally_2006,Skirvin_2024}, or the fast mode--\alfven{} wave conversion in the transition region layer \citep[][]{Cally_2011,Cally_2017}. This theory is supported by the coexistence of slow and fast PDs on the same structures, which could correspond to the signatures of the mother wave and of the converted wave, respectively.  Future statistical work is required to verify if fast and slow PDs are connected in such a way or not. 

Finally, contrary to previous observations of transverse waves in the corona \citep[][]{Morton_2015,Morton_2016}, fast PDs appear to mainly propagate in one direction, from the footpoint to the top of the coronal loops. If fast PDs were generated from both footpoints, and considering their little to no damping with distance, one would expect to see counter-propagative ones as well. The reason why this is not the case is still unclear. We suggest that fast PDs could be generated from a single footpoint only, possibly due to a specific magnetic configuration. The "single footpoint origin" scenario is supported by the non-detection of fast PDs in eight out of the 13 slits shown in Fig.\ref{fig:results:fov_time_distance_maps_rundiff_vel}a, b (the red slits). Thus, there are cases of loops (e.g., s$_2$) and loop footpoints (e.g., s$_4$) that show no clear signature of fast PDs. This indicates that in some cases, fast PDs might not be generated at both of the loop footpoints. Similarly, only specific footpoint configurations could produce fast PDs with a sufficiently high intensity increase above the background to be detectable by HRIEUV. Another scenario explaining the single direction propagation of fast PDs is the damping at high altitudes. This scenario would mainly apply to large coronal loops where the high altitude emitting plasma is not visible in HRIEUV (e.g., s$_1$ in Fig.\ref{fig:results:fov_time_distance_maps_rundiff_vel}a).

\section{Conclusion}

In this work, we have reported on small intensity enhancements along AR loops propagating with a wide range of PoS velocities reaching up to \SI{2200}{\kilo\meter\per\second}. These fast PDs are mainly visible in the upper part of the coronal loops, show little to no damping along their propagation path, and repeat in a two-minute periodic pattern. Their origin seems to be located near the footpoints of the coronal loops, but their exact nature remains unclear. Still, they show properties consistent with upflows, current sheets generated by propagating transverse oscillations, and fast MHD modes or torsional \alfven{} waves. They show potential as signatures of mechanisms able to transport the energy and mass from the photosphere or lower atmosphere to the upper layers of the Sun and as potential diagnostics of localized energy depositions at the footpoints of the coronal loops.

\begin{acknowledgements}
The authors would like to thank the referee for the suggestions that greatly helped to improve the manuscript. The authors would like to thank S. Parenti and G. Valori for coordinating the SOOP. The authors would also like to thank M. Janvier, J. Touresse and E. Pariat for the fruitful discussions. The work of A. Dolliou is funded by the Federal Ministry for Economic Affairs and Climate Action (BMWK) through the German Space Agency at DLR based on a decision of the German Bundestag (Funding code: 50OU2101, 50OU2201). This research was supported by the International Space Science Institute (ISSI) in Bern, through ISSI International Team project \#24-601 (Active region evolution under the spotlight, with unprecedented coordinated high-resolution stereoscopic observations and numerical simulations). Solar Orbiter is a space mission of international collaboration between ESA and NASA, operated by ESA. The EUI instrument was built by CSL, IAS, MPS, MSSL/UCL, PMOD/WRC, ROB, LCF/IO with funding from the Belgian Federal Science Policy Office (BELSPO/PRODEX PEA 4000112292 and 4000134088); the Centre National d’Etudes Spatiales (CNES); the UK Space Agency (UKSA); the Bundesministerium für Wirtschaft und Energie (BMWi) through the Deutsches Zentrum für Luft- und Raumfahrt (DLR); and the Swiss Space Office (SSO).  CF was supported by the Centre National d’Études Spatiales (CNES), through its APR program and by the Action Thématique Soleil-Terre (ATST) of CNRS/INSU PN Astro, co-funded by CNES and CEA. This work used data provided by the MEDOC data and operations centre (CNES / CNRS / Univ. Paris-Saclay), \url{http://medoc.ias.u-psud.fr/}. This research used version 6.1.1 \citep[][]{stuart_j_mumford_2025_14919826} of the SunPy open source software package \citep[][]{sunpy_community2020}. This research used version 0.10.1 \citep[][]{will_barnes_2025_14861973} of the aiapy open source software package \citep[][]{Barnes2020}. TVD received financial support from the Flemish Government under the long-term structural Methusalem funding program, project SOUL:
Stellar evolution in full glory, grant METH/24/012 at KU Leuven. The research that led to these results was subsidised by the Belgian Federal Science
Policy Office through the contract B2/223/P1/CLOSE-UP. It is also part of the DynaSun project and has thus received funding under the Horizon
Europe programme of the European Union under grant agreement (no. 101131534). Views and opinions expressed are however those of the author(s) only and
do not necessarily reflect those of the European Union and therefore the European Union cannot be held responsible for them. Portions of the numerical implementation in Appendix \ref{sec:annex:ponderomotive_force:python} were developed with assistance from ChatGPT (OpenAI, GPT-5.3), with subsequent verification by the authors.
\end{acknowledgements}

\bibliographystyle{aa}
\bibliography{Biblio.bib}

\begin{appendix}

\section{Checking that fast PDs are not caused by transverse oscillations of the coronal loops}
\label{sec:annex:verify_artifacts_transverse}

\begin{figure*}
    \includegraphics{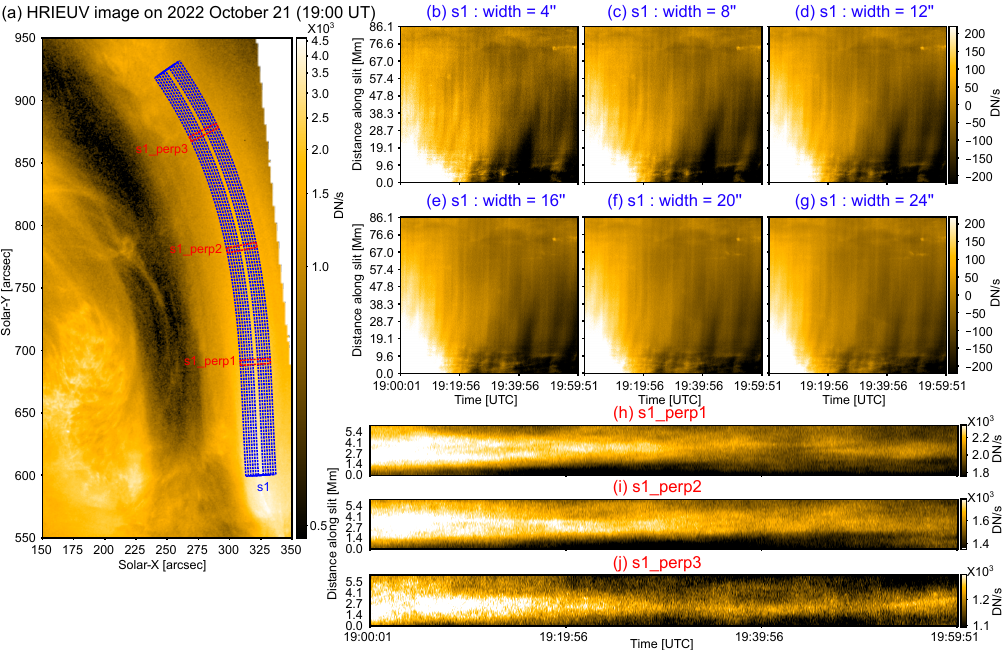}
    \caption{Testing that fast PDs detected along s$_1$ are not artifacts caused by transverse oscillations of the strands. (a) An HRIEUV image is zoomed-in around slit s$_1$, which is displayed with increasing widths from \SI{4}{\arcsec} to \SI{24}{\arcsec} (blue dotted lines). Three perpendicular cuts along the slits, called s1\_perp1 to s1\_perp3 are also displayed in red. The time distance maps computed along slit s$_1$ with an increasing width are displayed from (b) to (g). Each row is subtracted by their temporal mean over the whole sequence. The time distance maps of the perpendicular cuts (without subtraction of the temporal mean) are displayed from (h) to (j).}
    \label{fig:annex:checking_fast_waves_s1}
\end{figure*}

\begin{figure*}
    \includegraphics{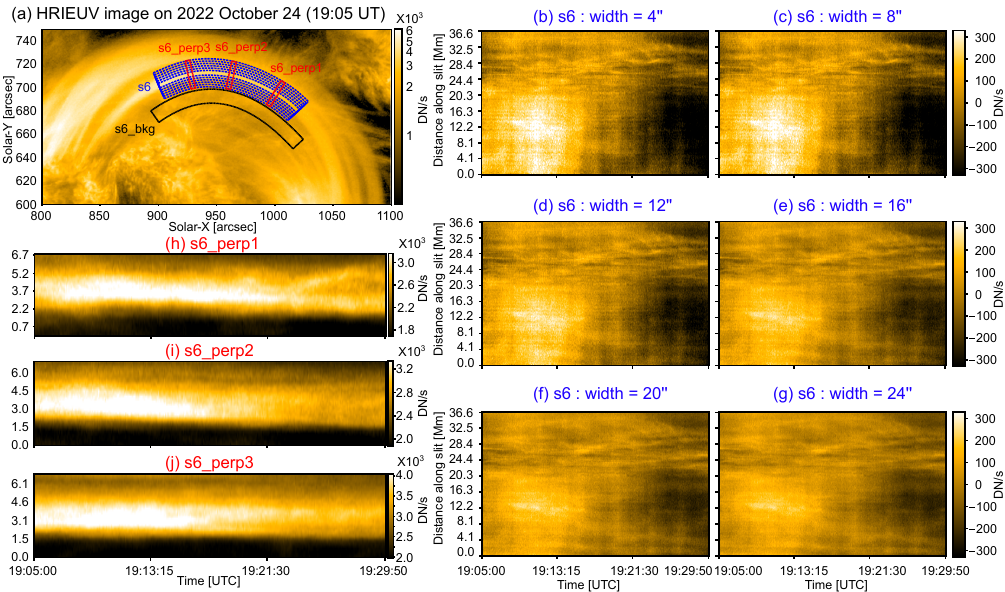}
    \caption{Same as for Fig.\ref{fig:annex:checking_fast_waves_s1}, but for the part of slit s$_6$ where fast PDs are detected. The black slit displayed in (a) corresponds to the "background" slit in Fig.\ref{fig:annex:time_distance_bckg}b.}
    \label{fig:annex:checking_fast_waves_s6}
\end{figure*}

\begin{figure}
    \centering
    \includegraphics{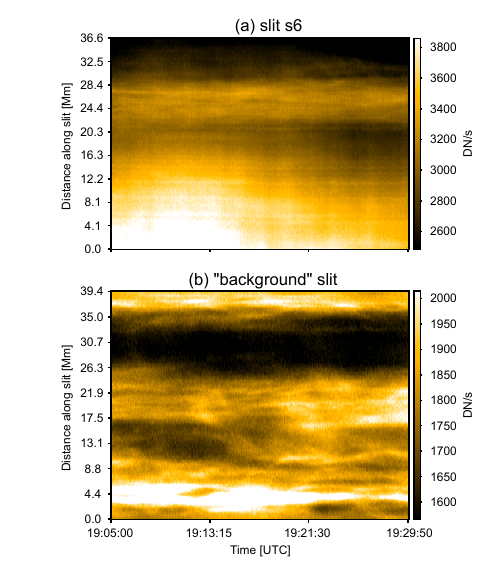}
    \caption{Zoomed-in portion of the time distance map for slit s$_6$ where fast PDs are detected (a), and of the background slit (b) displayed in black in Fig.\ref{fig:annex:checking_fast_waves_s6}a. }
    \label{fig:annex:time_distance_bckg}
\end{figure}

\begin{figure*}
    \centering
    \includegraphics{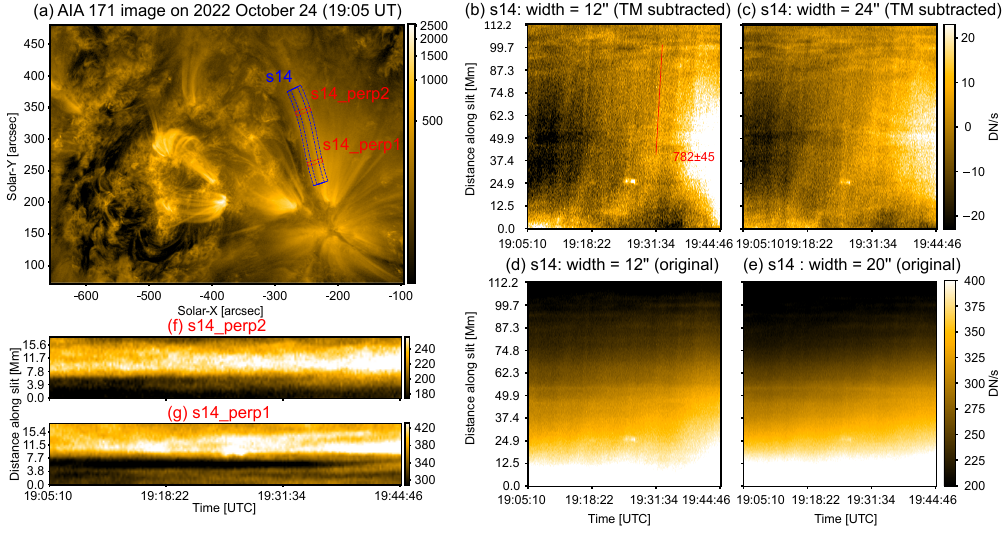}
    \caption{Same as Fig.\ref{fig:annex:checking_fast_waves_s1} on 2022 October 24 but for the AIA 171 dataset (19:00 to 19:45 UT). The time distance maps (b) to (e) corresponds to those obtained by slits of \SI{12}{\arcsec} and \SI{24}{\arcsec}. The time distance maps (b) and (c) have each of their row subtracted by the total temporal mean over the whole sequence, while (e) to (g) show the original data. The slit s$_{14}$ has been placed along a strand that at a location close to that of s$_{13}$ (Fig.\ref{fig:results:fov_time_distance_maps_rundiff_vel}b). "TM" stands for "temporal mean".} 
    \label{fig:annex:checking_fast_waves_s14}
\end{figure*}

In this section, we verify that the fast PDs are not artifacts caused by transverse oscillations of the strand moving in and out of the slits. Combined with the limited slit width, such oscillations could induce spurious intensity fluctuations in the time distance maps. We chose to perform two distinct tests, that we applied to the slit s$_1$ (Fig.~\ref{fig:annex:checking_fast_waves_s1}a) and to the top of slit s$_6$, where fast PDs are detected (Fig.~\ref{fig:annex:checking_fast_waves_s6}a).  

For the first test, we verified that the same fast PDs could be detected for a wide range of slit widths  (initially at \SI{12}{\arcsec}) from \SI{4}{\arcsec} to \SI{24}{\arcsec}. For each width, the time distance maps are computed along slit s$_1$ (Fig.~\ref{fig:annex:checking_fast_waves_s1}b to c) and s$_6$ (Fig.~\ref{fig:annex:checking_fast_waves_s6}b to c). The temporal mean over the whole time interval has been subtracted to each row, as for Fig.~\ref{fig:results:fov_time_distance_maps_rundiff_vel}. The same fast PDs are detected for every slit widths. This indicates that fast PDs are not the result of a limited slit width.

For the second test, we placed three slits perpendicular to the axis of s$_1$ (Fig.\ref{fig:annex:checking_fast_waves_s1}a) and s$_6$ (Fig.~\ref{fig:annex:checking_fast_waves_s6}a). The time distance maps of the perpendicular cuts are shown in Fig.~\ref{fig:annex:checking_fast_waves_s1}, \ref{fig:annex:checking_fast_waves_s6}h to j. No transverse oscillation that could move a bright strand outside of the slit were detected. We concluded from the results of these two tests that fast PDs detected in slits s$_1$ and s$_2$ are not caused by transverse oscillations that move the strand in and out of the slit. 

Furthermore, we verified that fast PDs detected on slit s$_6$ did not originate from the intensity fluctuation of the background or foreground along the LOS. To do so, we created a slit spatially shifted from the slit s$_6$ to estimate the background intensity outside of the coronal loop bundle (black slit in Fig.~\ref{fig:annex:checking_fast_waves_s6}a). The time distance maps show the fast PDs for slit s$_6$ (Fig.~\ref{fig:annex:time_distance_bckg}a), but not for background slit (Fig.~\ref{fig:annex:time_distance_bckg}b). We conclude that fast PDs detected along slit s$_6$ do not originate from background/foreground intensity fluctuations.

Finally, we checked that the fast PDs detected on AIA 171 in Fig.\ref{fig:results:aia171_fastpds_v3} are not artifacts caused by transverse oscillations of the strand. To do so, we performed the same two tests as those applied to the HRIEUV data (Fig.\ref{fig:annex:checking_fast_waves_s14}). The results show that fast PDs could still be detected when the width of the slit increases from \SI{12}{} to \SI{24}{\arcsec} (Fig.~\ref{fig:annex:checking_fast_waves_s14}b to e). Also, perpendicular cuts confirm the lack of strong transverse oscillations (Fig.~\ref{fig:annex:checking_fast_waves_s14}f, g). As such, we conclude that fast PDs detected with AIA 171 are not artifacts.

\section{Checking for instrumental artifacts}
\label{sec:annex:verify_artifacts_instrumental}

\begin{figure}
    \centering
    \includegraphics{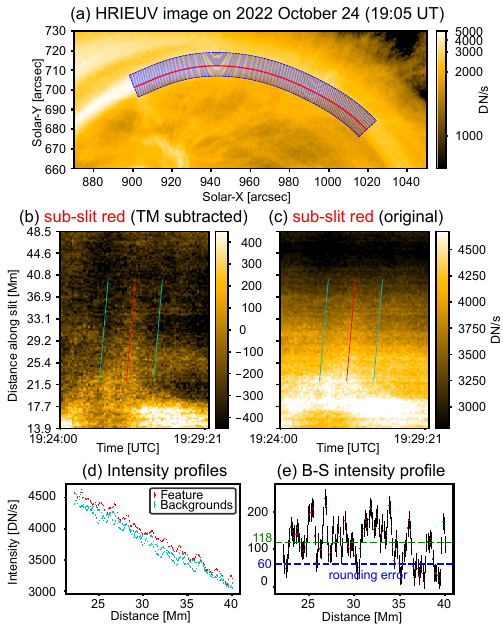}
    \caption{Impact of the recoding error on the detection of F$_{14}$. (a) HRIEUV Image showing the slit s$_6$ split into sub-slits of 1 pixel width (blue full lines). The red line indicates the location of the sub-slit where F$_{14}$ is detected. Sub-figures (b) to (e) are similar to those shown in Fig.\ref{fig:results:AR_21_10_22T1900_s1_damping_0} (TM stands for temporal mean). The time distance maps are computed along the sub-slit indicated as a red line in (a). The average over the distance (green) and the recoding error values (blue) are indicated as horizontal dotted lines in (e) }
    \label{fig:annex:check_rounding_s6}
\end{figure}

In this section, we review the instrumental artifacts that might induce fake detection of fast PDs (Section \ref{sec:annex:verify_artifacts_instrumental:detection_fastpd}) and fake detection of peaks in the Fourier power (Section \ref{sec:annex:verify_artifacts_instrumental:detection_period}). 

\subsection{Detection of fast PDs}
\label{sec:annex:verify_artifacts_instrumental:detection_fastpd}

Due to the low intensity peak of fast PDs above the background (Section \ref{sec:results:damping}), we checked that they are not caused by high gain (HG) - low gain (LG) switch ; compression effects ; limitations of the HRIEUV temporal and spatial resolution ; and recoding errors.

In order to obtain a compromise between saturation and readout noise, pixels in EUI are divided into HG pixels (low saturation level, low read noise) and LG pixels \citep[high saturation level, high read noise,][]{EUI_instrument}. The threshold HG-LG is at \SI{1118}{\dn} in L1 images and \SI{2409}{\dn\per\second} in L2 images (taking into account the exposure time of \SI{1.65}{\second} and the scaling factor of 3.555556 to HG values applied during L1 to L2 processing). The switch from HG to LG pixels can produce spurious bright or dark structures due to the sudden change in the readout noise. However, the fast PDs are mainly detected in either pure HG area, such as s$_1$ and s$_5$ (from \SI{2000}{} to \SI{1000}{\dn\per\second}), or pure LG area such as in s$_6$ (from \SI{3800}{} to \SI{3000}{\dn\per\second}). Therefore, fast PDs are not caused by a switch from HG to LG pixels.

The spatial resolution of HRIEUV is equal to 2 pixels \citep[][]{Berghmans_2023}. Due to the longitudinal extent of fast PDs (up to \SI{30}{\mega\meter}), they are not a consequence of the instrumental Point Spread Function (PSF, $\approx$ \SI{300}{\kilo\meter}). Similarly, compression artifacts are restrained to a few pixels, and cannot be responsible for fast PDs. Furthermore, the PoS velocities of fast PDs are all above the limit $v_\mathrm{cad}$ temporally resolvable by HRIEUV : $v_\mathrm{cad} = D/\mathrm{dt}$. Here, $\mathrm{dt}=$ \SI{5}{\second} is the HRIEUV cadence and $D$ is the distance over which the PoS velocity is measured. For instance, F$_1$ (\SI{2200}{\kilo\meter\per\second}) is characterized by a shift of \SI{15}{\second} (three HRIEUV cadence steps) over the distance of \SI{36}{\mega\meter}.

The low intensity peak values of fast PDs compared to the background intensity (Section \ref{sec:results:damping}) required to carefully evaluate the impact of recoding errors, which we estimated with the data count (DN) histograms over all pixels. Indeed, these histograms do not show a continuous distribution, but multiple peaks separated by values that increase with DN. The recoding error was estimated by taking half the distance between each of these peaks at the DN range of the different slits. This results in a maximum recoding error of \SI{35}{\dn\per\second} (slits s$_1$, s$_5$) and of \SI{60}{\dn\per\second} (slit $s_6$). These values cannot be directly compared to the \SI{10}{} to \SI{30}{\dn\per\second} intensity peaks computed in Section \ref{sec:results:damping}, as the time distance maps were obtained by spatially averaging the data over slit widths of \SI{12}{\arcsec} (24 pixels). This step necessarily decreases the intensity peak value of fast PDs compared to the background. To fully estimate the intensity peak value of fast PDs over a single pixel, we performed the test shown in Fig.~\ref{fig:annex:check_rounding_s6} to F$_{14}$ in s$_6$. We note that we also used the results from this method to estimate the intensity fluctuations of fast PDs compared to the background (see Section \ref{sec:results:damping}). First, we divided s$_6$ into sub-slits parallel to the slit axis and of 1 pixel width (Fig.~\ref{fig:annex:check_rounding_s6}a). We created time distance maps for each sub-list and we identified the location of F$_{14}$ (Fig.~\ref{fig:annex:check_rounding_s6}b, c). We then estimated the background-subtracted intensity profile (Fig.~\ref{fig:annex:check_rounding_s6}d, e), using a similar method as the one described in Section \ref{sec:results:damping}. With regards to F$_{14}$, the background-subtracted intensity profile shows peaks up to \SI{230}{\dn\per\second} and an average of \SI{118}{\dn\per\second} over the total distance, which is above the \SI{60}{\dn\per\second} recoding error at \SI{4600}{\dn\per\second}. As such, F$_{14}$ is not caused by recoding errors. Applying the same test to F$_{12}$ and F$_{13}$, we found averages over distance of \SI{110}{\dn\per\second} and \SI{82}{\dn\per\second} and peaks up to \SI{250}{\dn\per\second} and \SI{200}{\dn\per\second}, respectively. We conclude that fast PDs detected along s$_6$ are not caused by recoding errors.  

We applied a similar test to the other fast PDs detected along the slits s$_1$ and s$_5$. To stay consistent in term of distance to the test done in s$_6$, we restrained the analysis to \SI{15}{\mega\meter} around the intensity peak of the fast PDs. Five of the background-subtracted intensity peak profiles of fast PDs have an average above the recoding error of \SI{37}{\dn\per\second}, ranging from \SI{58}{\dn\per\second} (F$_{9}$) to \SI{42}{\dn\per\second} (F$_8$), with peaks up to \SI{130}{\dn\per\second}. Two of them (F$_{10}$, F$_{11}$) have an average close to the detection limit, from \SI{37}{} to \SI{31}{\dn\per\second}, with peaks up to \SI{100}{\dn\per\second}. Two of the fast PDs (F$_4$ and F$_5$) could not be detected with the method described in the last paragraph. This was to be expected, as transverse oscillations within the slit width are likely to diffuse the intensity associated with fast PDs over multiple sub slits. The fast PD would then be barely detectable above the background on an individual sub slit. We note that all fast PDs have an intensity fluctuation of 2 to 4\% of the total intensity, when computed on 1 pixel width slits (See Section \ref{sec:results:damping}). 

Finally, while still being close to the detection limit, we conclude that fast PDs are not caused by recoding error, and instrumental artifacts in general. We provide the following arguments : (1) most of fast PDs had an intensity profile with an average over distance higher than the recoding error level, up to double the value for F$_{12}$ and F$_{14}$ ; (2) recoding errors, and instrumental artifacts in general, are unlikely to produce intensity peak motions with a temporally resolved PoS velocity along specific strands, especially given the wide range of intensity values (e.g., from \SI{4600}{} to \SI{3000}{\dn\per\second} for slit s$_6$) ; (3) fast PDs are also detected with AIA 171 (Fig.\ref{fig:results:aia171_fastpds_v3}), which has a negligible recoding error of \SI{0.5}{\dn\per\second} in L1.5 FITS files. 

\subsection{Detection of the \SI{2}{\min} peak in the Fourier power}
\label{sec:annex:verify_artifacts_instrumental:detection_period}

In section \ref{sec:results:periodicity}, we measured significant peaks at \SI{2}{\min} on the Fourier powers associated with fast PDs in slit s$_6$. In this section, we review potential instrumental artifacts that could induce fake peak detection the Fourier power. 

Fixed patterns on the detector, such as Flat Field (FF), can induce fake periodic pattern after the reprojection to Carrington coordinates \citep[][]{Auchere_2014}. Regarding HRIEUV, patterns on the detector include : the FF ; a bias correction every four columns applied during data processing \citep[][]{EUI_instrument} ; and a "fixed noise pattern" recently discovered in the data release 6.0. In our case however, the observation was set in a "tracking" mode. As such, the pointing of the satellite compensated for the Carrington rotation of the Sun. Reprojecting the images into Carrington coordinates should then result in minimal artifacts due to fixed patterns on the detector. 

However, the correction of the jitter involves the reprojection of images into new grids, which can also generate spurious periods from the fixed patterns on the detector. To quantify this effect, we measured the evolution with time of the "CRVAL1" values on the FITS header, that oscillates due to the jitter correction. A FFT applied to this time series showed a peak in the Fourier power at a period of \SI{50}{\second} ($2\cdot 10^{-2}$ Hz). Spurious peaks associated with this frequency can be found all along the slit in Fig.\ref{fig:results:AR24_s6_periodicities}. However, since this spurious frequency is higher that that of the measured signal ($7\cdot 10^{-3}$ Hz), this latter can't be a harmonic of the jitter frequencies.

\section{Method to compute Fourier transform and the 95\% confidence level}
\label{sec:annex:FFT}

In this section, we describe the method used to compute the Fourier power of the data and of the background, as well as the 95\% confidence level in Fig.\ref{fig:results:AR24_s6_periodicities}. The processing itself is similar to the one used in \citep{Froment_2015,Auchere_2014}. First, the light curves were subtracted by their temporal mean over the whole time interval. No detrending was applied to the light curve, as it can produce spurious frequencies in the Fourier spectrum \citep[][]{Auchere2016}. The light curve was then divided by its variance and an apodization was performed with a Hanning window. The Fourier power was obtained with a fast Fourier transform algorithm (FFT). 

We normalized the Fourier power by the variance of the light curve \citep{Gabriel_2002,Froment_2015,Auchere2016}. Similarly to \citet{Kayshap_2019} and \citet{Auchere2016}, we chose to model the Fourier power of the background $\mathrm{P}_\mathrm{bkg}(\nu)$ with a power law  model: 

\begin{equation}
\label{eq:P_bkg}
    \mathrm{P}_\mathrm{bkg}(\nu) = \mathrm{A}\nu^\mathrm{s} + \mathrm{C}.
\end{equation}

The parameters (A, s, c) were obtained by applying a least-square fit to the Fourier power of the data. We defined the 95\% confidence limit as $m$ times $\mathrm{P}_\mathrm{bkg}(\nu)$. The constant $m$ was set to $m=-\log\left(1 - 0.95^{2/N}\right)$, with $N$ being the total number of time steps in the time series \citep[][]{Gabriel_2002,Auchere_2016}. With this definition, there was 5\% probability for at least one frequency to peak above the 95\% confidence level by chance. In Fig.~\ref{fig:results:AR24_s6_periodicities}, we obtained $m=7.97$ for $N=296$.

\section{Density perturbations induced by the ponderomotive force in Alfv\'en waves}
\label{sec:annex:ponderomotive_force}

In this section, we show that density perturbation induced by the ponderomotive force in Alfv\'en waves are a possible model to explain fast PDs. In Section \ref{sec:annex:ponderomotive_force:analytical}, we derive the velocity amplitude of the mother wave that could induce density perturbation consistent with the intensity fluctuation we measured in HRIEUV (4\% to 8\%). We confirm our analytical derivation with 1.5 dimensions (1.5D) ideal MHD simulations in Section \ref{sec:annex:ponderomotive_force:python}.

\subsection{Analytical derivation}
\label{sec:annex:ponderomotive_force:analytical}

We consider an equilibrium magnetic field aligned with the $z$-direction: $\vec{B}=B_0\vec{e}_z$, in a homogeneous plasma with density $\rho_0$. In 
this, a periodic \alfven{} wave propagates with wave components $\vec{B}_1=B_{1y}\vec{e}_y$ and $\vec{V}_1=V_{1y}\vec{e}_y$. We used the following equation for the induced density $\rho_2$, as also derived by \citet{Hollweg_1971,Rankin_1994} and used by \citet{terradas2004}:
\begin{equation}
	\frac{\partial^2 \rho_2}{\partial t^2}-v_\mathrm{s}^2 \frac{\partial^2\rho_2}{\partial z^2} = \frac{1}{2\mu} \frac{\partial^2 B_{1y}^2}{\partial 
z^2},
	\label{eq:noncold}
\end{equation}
where $\mu$ is the magnetic permeability and $v_\mathrm{s}$ refers to the sound speed, respectively. The RHS in this equation is the ponderomotive force leading to compressions.

We consider the mother \alfven{} wave to have the form of a general pulse, which we denote as

\begin{equation}
	B_{1y} = \delta B\ {\cal G}\left(\frac{z-v_\mathrm{A}t}{\sigma_z}\right),
\end{equation}
where ${\cal G}(\phi)$ is a function of a single variable $\phi$, and $\phi=\frac{z-v_\mathrm{A}t}{\sigma_z}$ is the phase of the function, with $v_\mathrm{A}$ the \alfven{} speed and $\sigma_z$ a scale parameter along $z$. The phase reflects the propagating nature of the mother \alfven{} wave. Similar to d'Alembert's solutions to 
the wave equation, the only mathematical condition for the ${\cal G}$ function is to be twice differentiable. We chose to normalize 
$\max{({\cal G})}=1$ in order to keep all amplitude dependence in the $\delta B$ parameter. We may calculate the derivative of ${\cal G}$ with the chain rule:
\begin{equation}
\frac{\partial}{\partial t} {\cal G}(\phi) = {\cal G}' \frac{\partial \phi}{\partial t} = -\frac{v_\mathrm{A}}{\sigma_z}{\cal G}'. 
\end{equation}
When we also assume that the induced density perturbation $\rho_2$ is only a function of $\phi$, we find from Eq.~\ref{eq:noncold} that we have the following as a
solution:
\begin{equation}
	\rho_2 = \frac{1}{2\mu} \frac{\delta B^2}{v_\mathrm{A}^2-v_\mathrm{s}^2} {\cal G}^2\left(\frac{z-v_\mathrm{A}t}{\sigma_z}\right).
\end{equation}
Further rewriting it to normalized quantities yields
\begin{equation}
	\frac{\rho_2}{\rho_0} = \frac{1}{2} \frac{\delta B^2}{B_0^2}\frac{1}{1-\frac{v_\mathrm{s}^2}{v_\mathrm{A}^2}} {\cal 
G}^2\left(\frac{z-v_\mathrm{A}t}{\sigma_z}\right).
\end{equation}
This equation shows that the density perturbation is co-propagating with the mother wave pulse. In case of $v_\mathrm{s}\ll 
v_\mathrm{A}$, the amplitude of the induced density perturbation is solely determined by the square of the amplitude of the mother wave. For 
comparison with observations, it is often easiest to rewrite it in terms of the velocity amplitude of the mother wave:
\begin{equation}
	\frac{\rho_2}{\rho_0} = \frac{1}{2} \frac{\delta v^2}{v_\mathrm{A}^2}\frac{1}{1-\frac{v_\mathrm{s}^2}{v_\mathrm{A}^2}} {\cal 
G}^2\left(\frac{z-v_\mathrm{A}t}{\sigma_z}\right),
\label{eq:densvel}
\end{equation}
where $\delta v$ is the velocity amplitude of $V_{1y}$. 

We now apply this to the observational results from Section \ref{sec:results:damping}. We found an intensity fluctuation ($\delta I$) compared to the total intensity ($I_0$) of $\delta I/I_0\sim 4-8\%$ in HRIEUV for fast PDs. Assuming the intensity to scale with density squared $I\sim n^2$, we may postulate that the intensity variations will be twice the density variations:
\begin{equation}
	\frac{\delta I}{I_0} = 2 \frac{\delta \rho}{\rho_0}.
\end{equation}
Using Eq.~\ref{eq:densvel}, we then find that the velocity perturbations of the mother wave must have a relative velocity amplitude in the range 
$\frac{\delta v}{v_\mathrm{A}}=\sqrt{0.04}-\sqrt{0.08} = 0.20-0.28$. For an observed propagation speed between 500 and \SI{2200}{\kilo\meter\per\second}, we thus find velocity 
amplitudes of the mother wave of \SI{440}{} to \SI{620}{\kilo\meter\per\second} to generate the observed induced density variations. 

The result in Eq.~\ref{eq:densvel} also holds for multi-dimensional simulations. If we consider ${\cal G}$ to also have a spatial dependence,
\[ {\cal G}(x,y,z-v_\mathrm{A}t), \]
then all the above computations hold true. The induced density perturbation will also then scale as follows:
\begin{equation}
	\frac{\rho_2 (x,y,\phi)}{\rho_0} = \frac{1}{2} \frac{\delta B^2}{B_0^2}\frac{1}{1-\frac{v_\mathrm{s}^2}{v_\mathrm{A}^2}} {\cal 
G}^2\left(x,y,z-v_\mathrm{A}t\right).
\end{equation}
This allow the density variations simulated by \citet{Shestov_2017} to be predicted.

\subsection{1.5D ideal MHD simulations}
\label{sec:annex:ponderomotive_force:python}

We used an ideal MHD model in 1D with two velocity and magnetic field components (so-called 1.5D). The following set of differential equations is solved:
\begin{align}
	\frac{\partial \rho}{\partial t} & = - \frac{\partial}{\partial x}(\rho v_x) + \nu \frac{\partial^2}{\partial x^2} \rho, \\
	\frac{\partial v_x}{\partial t} & = -v_x \frac{\partial v_x}{\partial x} - \frac{1}{\rho} \frac{\partial P}{\partial x} + \nu 
\frac{\partial^2}{\partial x^2} v_x, \\
	\frac{\partial v_y}{\partial t} & = -v_x\frac{\partial v_y}{\partial x} + \frac{1}{\mu\rho} B_x \frac{\partial B_y}{\partial x} + \nu 
\frac{\partial ^2}{\partial x^2} v_y, \\
	\frac{\partial B_x}{\partial t} & = 0, \\
	\frac{\partial B_y}{\partial t} & = -\frac{\partial}{\partial x}(v_xB_y) + B_x\frac{\partial}{\partial x}v_y + \nu 
\frac{\partial ^2}{\partial x^2} B_y, \\
	P & = v_\mathrm{s}^2\rho+\frac{B_x^2+B_y^2}{2\mu}.
\end{align}
These equations are solved with a leapfrogging scheme, with intermediate values calculated at $\delta t/2$ that are then used to compute the solution 
at $\delta t$. The parameter $\nu=10^{-5}$ is the numerical diffusivity, which is incorporated to ensure numerical stability. Central differencing is 
used for spatial derivatives. The time step is dynamically varied with a CFL condition:
\[ \delta t = 0.3 \frac{\delta x}{\max{(v_x+v_\mathrm{s}+v_\mathrm{A})}}. \]

The simulation is initialized with a uniform $B_x = 1$, $v_x=0$ and density $\rho=1$ in a simulation domain $x\in [0,1]$. The permittivity is also 
set to $\mu=1$. To initialize the \alfven{} wave pulse, we have set 
\begin{equation}
	B_y = -v_y= 0.2 \sin{\left(20\pi (x-0.0533)\right)}
\end{equation}
for $x \in [0.0533,0.0833]$, and 0 elsewhere. 

The results are displayed in Fig.~\ref{fig:pythonsims}. The mother \alfven{} wave is shown in the right panel, which shows that the initial 
perturbation in $v_y$ is propagating to the right with the \alfven{} speed. The left panel shows that the same location of the propagating mother 
\alfven{} wave pulse also has an associated density perturbation, which is also co-propagating with the \alfven{} speed. This numerical model 
confirms the analytical derivation in Eq.~\ref{eq:densvel}.

\section{Additional table and figures}
\label{sec:annex:additional_figures}

\begin{figure*}
    \centering
    \includegraphics[]{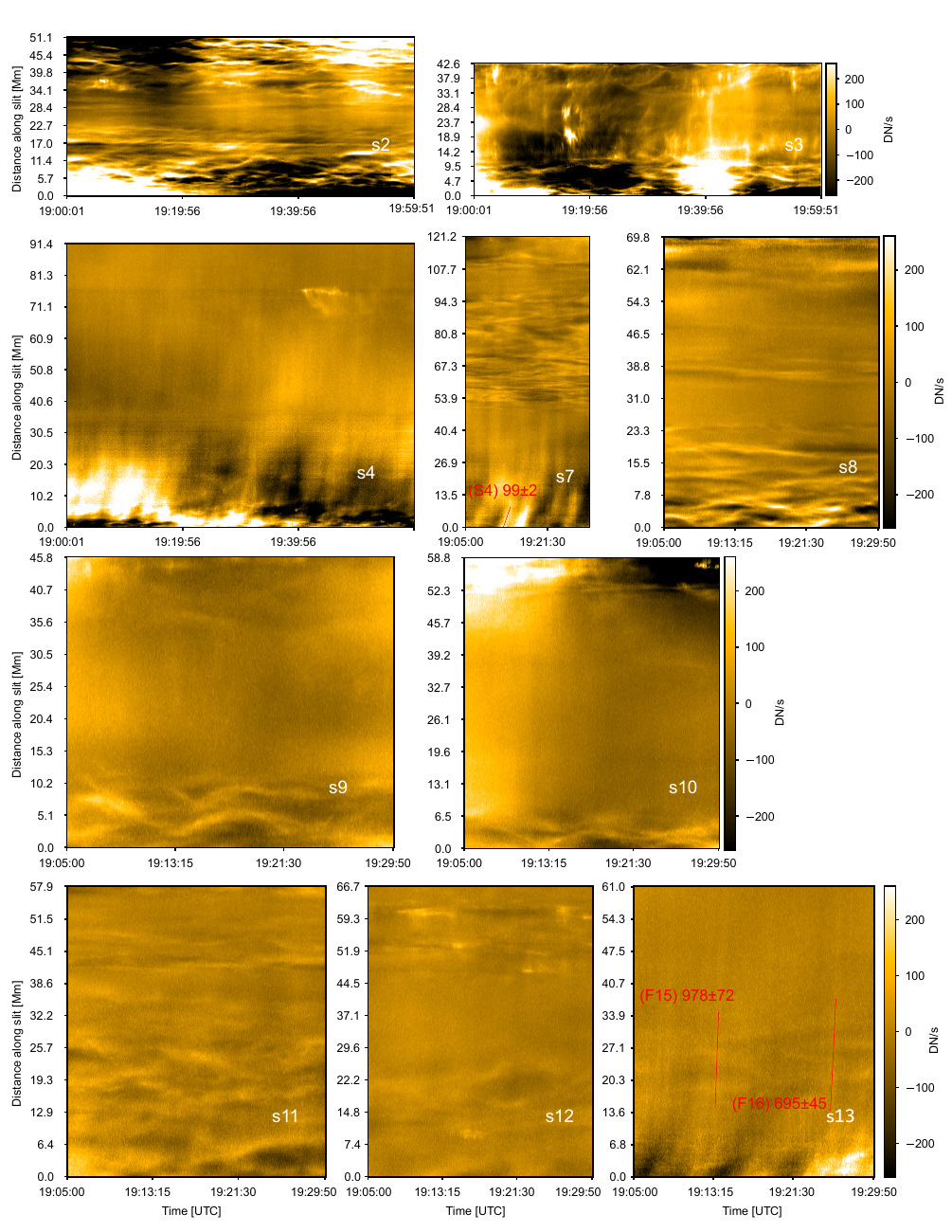}
    \caption{Time distance maps of all the slits other than s$_1$, s$_5$, s$_6$ for the sequences on 2022 October 21 and 24, similar to Fig.\ref{fig:results:fov_time_distance_maps_rundiff_vel}c,d,e. The label of each slit is indicated in the lower right for each sub-figure. The norm is the same for all sub-figures. }
    \label{fig:annex:AR21_AR24_all_other_slits_time_distance_v4}
\end{figure*}

\begin{figure*}
    \centering
    \includegraphics{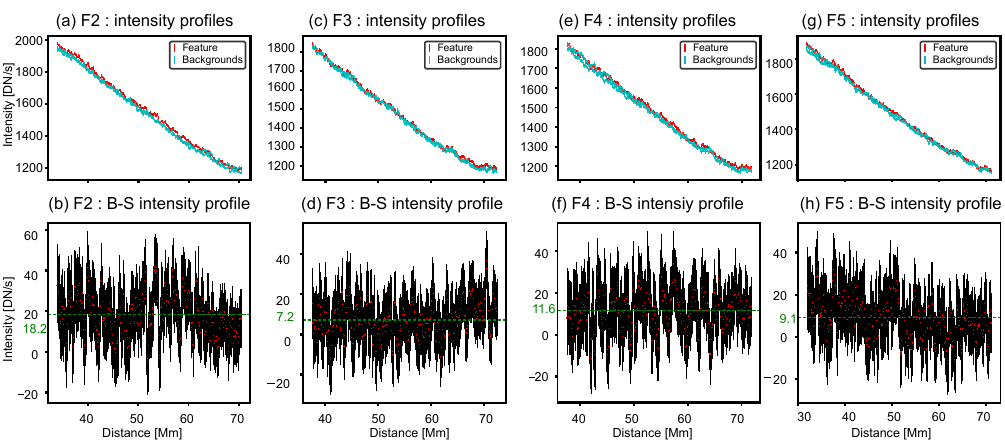}
    \caption{Damping analysis similar to Fig.\ref{fig:results:AR_21_10_22T1900_s1_damping_0}c, d, but for the fast PDs detected in s$_1$ other than F$_1$. "B-S" stands for "background-subtracted".  }
    \label{fig:annex:damping_all_others_s1}
\end{figure*}

\begin{figure*}
    \centering
    \includegraphics{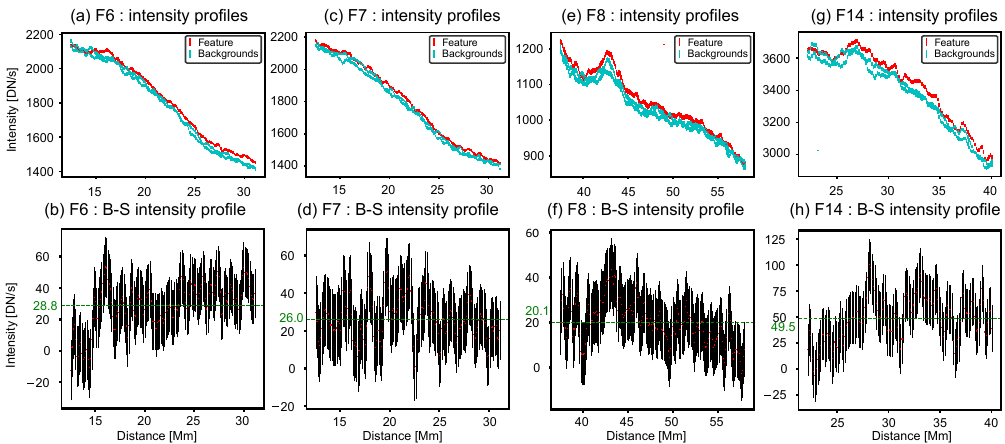}
    \caption{Damping analysis similar to Fig.\ref{fig:results:AR_21_10_22T1900_s1_damping_0}c, d, but for fast PDs detected in s$_5$ and s$_6$. "B-S" stands for "background-subtracted".}
    \label{fig:annex:damping_s5_s6}
\end{figure*}

\begin{figure*}
    \centering
    \includegraphics{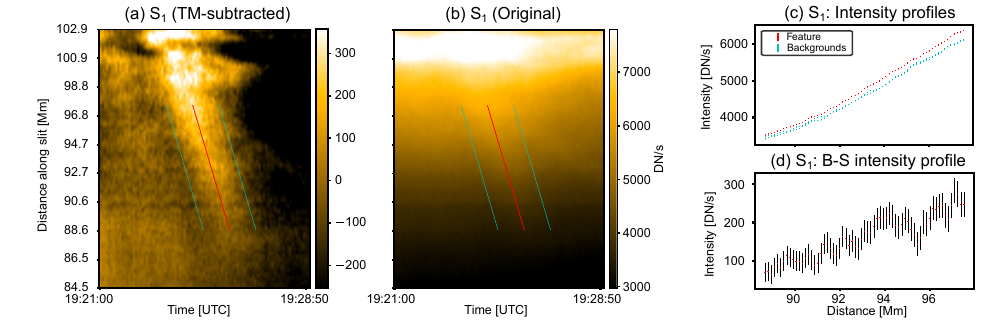}
    \caption{Same as for Fig.\ref{fig:results:AR_21_10_22T1900_s1_damping_0}, but for S$_3$ detected on slit s$_6$. "TM" and "B-S" respectively stand for "temporal mean" and "background subtracted."}
    \label{fig:results:AR_21_10_22T1900_s2_damping_2}
\end{figure*}

In this section, we propose a short description of the additional table and figures that are referred in the main text. The time distance maps along the slits s$_1$ to s$_{13}$, other than s$_1$, s$_5$ and s$_6$ (shown in Fig.\ref{fig:results:fov_time_distance_maps_rundiff_vel}), are given in Fig.\ \ref{fig:annex:AR21_AR24_all_other_slits_time_distance_v4}, while the PoS velocity values of all PDs are given in Table \ref{table:events}. Both table and figure are referred in Section \ref{sec:results:pos_velocity}. A damping analysis similar to the one applied in Fig.\  \ref{fig:results:AR_21_10_22T1900_s1_damping_0} is applied to fast PDs of s$_1$, s$_5$, and s$_6$ in Fig.\ \ref{fig:annex:damping_all_others_s1} and \ref{fig:annex:damping_s5_s6}. A damping analysis applied to the sloz PD S$_1$ is displayed in Fig.\ \ref{fig:results:AR_21_10_22T1900_s2_damping_2}. All three figures are referred in Section \ref{sec:results:damping}.

\renewcommand{\arraystretch}{1.5} 
\begin{table}
\caption{PoS velocities estimated for the fast and slow PDs }    
\flushleft          
\begin{tabular}{l c c c} 
\hline
\hline
Name & slit & $v_\mathrm{PoS}$ [\SI{}{\kilo\meter\per\second}] & $\mathrm{d}v_\mathrm{PoS}$ [\SI{}{\kilo\meter\per\second}] \\
\hline
F$_1$ & $s_1$ & 2204 & 201 \\
F$_2$ & $s_1$ & 1413 & 94 \\
F$_3$ & $s_1$ & 1440 & 86 \\
F$_4$ & $s_1$ & 1053 & 49 \\
F$_5$ & $s_1$ & 1100 & 43 \\
F$_6$ & $s_5$ & 953 & 70 \\
F$_7$ & $s_5$ & 1254 & 131 \\
F$_8$ & $s_5$ & 532 & 24 \\
F$_9$ & $s_5$ & 531 & 20 \\
F$_{10}$ & $s_5$ & 423 & 9 \\
F$_{11}$ & $s_5$ & 637 & 74 \\
F$_{12}$ & $s_6$ & 1697 & 372 \\
F$_{13}$ & $s_6$ & 1874 & 303 \\
F$_{14}$ & $s_6$ & 930 & 75 \\
F$_{15}$ & $s_{13}$ & 978 & 72 \\
F$_{16}$ & $s_{13}$ & 695 & 45 \\
\hline
\hline 
S$_{1}$ & $s_5$ & 86 & 2 \\
S$_{2}$ & $s_5$ & 82 & 3 \\
S$_{3}$ & $s_6$ & 105 & 1 \\
S$_{4}$ & $s_7$ & 99 & 2 \\

\end{tabular}      
\tablefoot{PoS velocity values ($v_\mathrm{PoS}$) and uncertainties ($\mathrm{d}v_\mathrm{PoS}$) estimated for the fast and slow PDs. For each PD, we indicate  its label and the slit along which is was detected.}
\label{table:events}  
\end{table}

\end{appendix}

\end{document}